\documentclass[twocolumn,showpacs,preprintnumbers,amsmath,amssymb,aps,prl]{revtex4}
\usepackage{graphicx}
\begin{document}

\title{Jamming in Systems  With Quenched Disorder }   
\author{C.J. Olson Reichhardt$^1$,    
E. Groopman$^{1,2}$, Z. Nussinov$^2$,
and C. Reichhardt$^1$}
\affiliation{
$^1$Theoretical Division,
Los Alamos National Laboratory, Los Alamos, New Mexico 87545, USA\\ 
$^2$Department of Physics, Washington University, St. Louis, Missouri
63160, USA} 

\date{\today}
\begin{abstract}
We numerically study the effect of adding quenched disorder in the
form of randomly placed pinning sites on jamming transitions 
in systems that jam at a well defined point J in the clean limit.
Quenched disorder 
decreases the  
jamming density and introduces a depinning threshold. 
The onset of a finite threshold coincides with point J 
at the lowest pinning densities, but for higher
pinning densities there is always a finite threshold even
well below jamming.
We find that
proximity to point J strongly affects the transport 
curves and noise fluctuations, and observe
a change from plastic behavior below jamming, where the system is highly
heterogeneous, to elastic depinning above jamming.  
Many of the general features we find are related  
to other systems containing quenched disorder, 
including the peak effect observed in vortex systems.
\end{abstract}
\maketitle

\vskip2pc
When a collection of particles such as grains is at low densities
with little grain-grain contact, the system acts like a liquid 
in response to an external drive.  At higher densities, however, significant
grain-grain contacts occur and the system responds like a rigid solid,
exhibiting a jamming transition where the grains can become stuck under an
external drive \cite{Nagel,L}.
The jamming density is termed point J in simple systems such as
bidisperse disk assemblies \cite{Nagel,OHern,L,N}.
An already jammed system can be unjammed by the application of shear
\cite{Teitel,B,Barret} and numerous studies have focused on
understanding jamming for varied grain shapes 
\cite{T}, interactions \cite{P}, temperatures \cite{Y},
and external drives \cite{Hastings,Dauchot}.  
Depinning is another example of a transition from a stuck or pinned state to
a flowing state under an applied drive, and occurs for collectively interacting
particles in quenched disorder such as
vortices in 
type-II superconductors \cite{Higgins,Marchetti,Olson}, 
colloids interacting 
with random or periodic substrates \cite{Reichhardt,Ling,Bechinger}, and
charge-density waves \cite{Gruner}.  
Above depinning, the particles pass from a stationary solid state into either
a flowing solid or a fluctuating, liquidlike state 
\cite{Higgins,Olson,Reichhardt}. 
Understanding 
how quenched disorder affects jamming and how jamming-unjamming transitions
are related to depinning would have a great impact in both fields.  

The first proposed
jamming phase diagram for loose particle assemblies had three axes: 
inverse density, load, and temperature \cite{Nagel}.
Here we propose that quenched disorder can form a fourth axis of the
jamming phase diagram,
and show that if a system has a well defined jamming 
transition in the absence of quenched disorder,
proximity to point J is relevant even for strong quenched disorder. 
We also show that jammed or pinned states below point J 
show profoundly different behaviors in response to an external drive 
compared to states above point J. 
We find that for varied amounts of disorder,  this
system exhibits many features found in vortex matter 
\cite{Higgins,Marley,Banerjee} 
including a peak effect near point J, suggesting that jamming
may be a useful way to understand many of the phenomena 
found in systems with pinning. 
We study a two-dimensional (2D) bidisperse disk
system with equal numbers of disks of each size that is
known to exhibit jamming at a well defined 
density for zero temperature and load.
The density is defined as the 
fraction $\phi$ of the system area that is covered by the disks.
For a disk radius ratio of $1:1.4$, in 2D the onset of jamming occurs 
at $\phi_{J} \approx 0.844$ \cite{OHern,Hastings,OlsonA,Teitel}.
Since point J in this system is well defined in the absence of quenched
disorder, we can determine how jamming changes when we add a small amount
of quenched disorder in the form of randomly placed pinning sites.
We also focus on distinguishing the
effect of jamming from that of depinning.
This
is particularly important since even 
non-interacting particles that do not exhibit a jamming transition 
in the absence of quenched disorder can still
exhibit a finite depinning threshold in the presence of disorder.

{\it Simulation--}
We consider a 2D system of size $L \times L$ 
with periodic boundary conditions in the $x$ and $y$-directions
containing 
$N$ disks that interact via a short range repulsive spring force. 
The sample is a 50:50 mixture of disks with interaction ranges $r_{A}$ and
$r_{B}$, where $r_{A}=1.4r_{B}$.
$N_J$ is the number of particles in the sample at the jamming transition.
To initialize the system, we place disks in nonoverlapping positions and then
shrink all disks, add a few additional disks, and reexpand all disks while
thermally agitating the disks until reaching the desired density.
We employ overdamped dynamics where the equation of motion for the disks is 
$ 
\eta d {\bf R}_{i}/dt = \sum_{i\neq j}k(R_{\rm eff}^{ij} - |{\bf r}_{ij}|)({\bf r}_{ij}/|{\bf r}_{ij}|)\Theta(R_{\rm eff}^{ij} - |{\bf r}_{ij}|) + {\bf F}_{p}^{i} + {\bf F}_{D}.
$  
Here the damping constant $\eta = 1$,
$k = 200$, and $R_{eff}^{ij} = r_{i} + r_{j}$, where $r_{i(j)}$ is the radius of
disk $i(j)$.     
The driving force ${\bf F}_{D} = F_{D}{\bf {\hat x}}$ is 
applied to all the disks uniformly.
The pinning force ${\bf F}_p^{i}$ 
is modeled as arising from $N_{p}$ non-overlapping attractive parabolic 
traps with a maximum force of $F_{p}$ and a cutoff radius of $r_{A}/2$ to ensure
that only one grain can be trapped by any one pinning site. 
In the absence of other grains,
an isolated grain will depin when $F_{D} > F_{p}$. 
In the absence of pinning, 
this model has been well studied and has a
jamming density of
$\phi_{J} \approx 0.844$ \cite{OHern,Hastings}. To determine if the 
system is pinned, we measure the total average grain velocity 
$\langle V_x\rangle =\sum_{i=1}^{N}{\bf v}_i \cdot {\bf \hat x}$ 
as a function of $F_{D}$, where ${\bf v}_i$ is the velocity of
grain $i$. For most of the results
presented, $F_D$ is slowly increased in
increments of $\delta F_D=5 \times 10^{-6}$ and we wait 
$5 \times 10^4$ simulation time steps after each force increment to ensure 
that the system reaches a steady state response.

\begin{figure}
\includegraphics[width=3.5in]{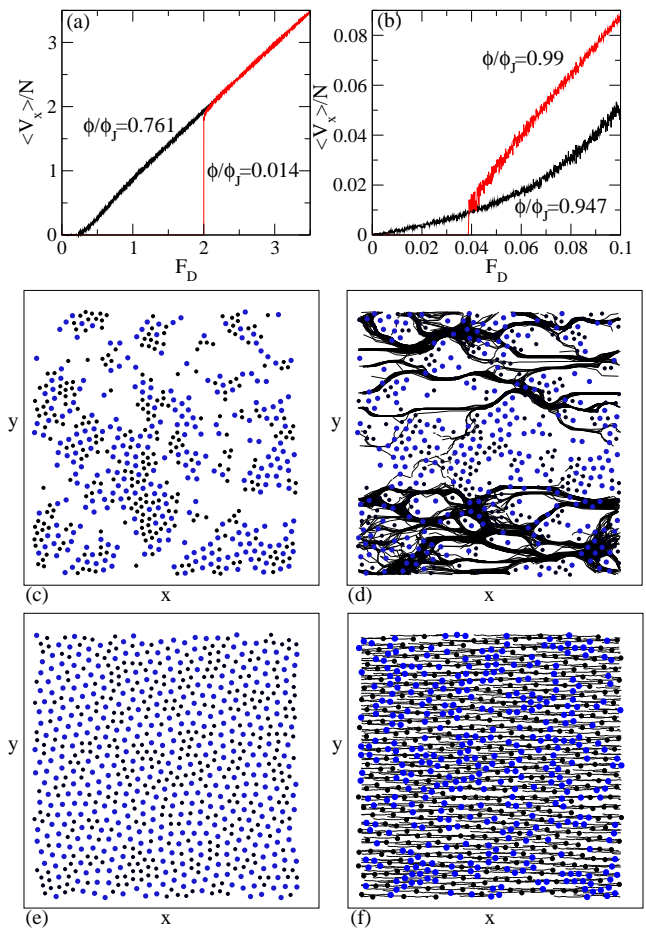}
\caption{ 
(Color online) (a,b) 
The normalized average grain velocity $\langle V_{x}\rangle/N$ 
vs external drive $F_{D}$ in samples with $F_p=2.0$. 
(a) $N_{p}/N_{J} = 0.415$. 
Right (red) curve: at $\phi/\phi_{J} = 0.014$ 
depinning occurs at $F_{c} = 2.0$. 
Left (black) curve: at $\phi/\phi_{J} = 0.761$, $F_{c}$ is much lower.
(b) $N_{p}/N_{J} = 0.09267$. 
Lower (black) curve: at $\phi/\phi_{J} = 0.947$, there is no finite depinning
threshold. 
Upper (red) curve: at $\phi/\phi_{J} = 0.99$, $F_{c}$ is finite. 
(c) Disk positions in the pinned state for the system in (a) with 
$\phi/\phi_{J} = 0.761$. 
(d) The disk trajectories over a period of time for the
system in (c) at $F_D=0.25$ showing plastic flow above the
depinning threshold.
(e) Disk positions in the pinned state for the system in (c) with
$\phi/\phi_J=1.03878$.
(f) The disk trajectories over a period of time for the system in
(e) at $F_D=1.1F_c$ showing elastic flow.
}   
\label{fig:1}
\end{figure}

In Fig.~1(a,b) we plot $\langle V_x\rangle/N$ versus $F_D$ 
for different pinning densities and particle densities, measured in terms of
$N_J$ and $\phi_J$.
At $N_{p}/N_{J} = 0.415$
and $\phi/\phi_{J}  = 0.014$ in Fig.~1(a), the system is pinned 
for drives up to $F_{D} = 2.0$.  At this low density, 
there are no interactions between the disks and the 
depinning threshold $F_c$ is solely determined by the pinning force.
For the same pinning density at $\phi/\phi_{J} = 0.761$, 
there are more disks than pinning sites; however, as shown in Fig.~1(a),
there is still a finite depinning threshold of $F_{c} = 0.27$. 
Since not all of the disks can be captured by pinning sites, the existence of
a finite depinning threshold indicates that the disks trapped by pins must be
blocking the flow of the disks that are not in pins.  Thus, 
in pinned states such as that illustrated in Fig.~1(c), 
some form of jamming must be occurring.
The depinning threshold can vanish
when the pinning density is reduced,
as shown in Fig.~1(b) for 
a sample with $N_{p}/N_{J} = 0.09267$ and $\phi/\phi_{J} = 0.947$. 
Here $F_c=0$ and the velocity response is nonlinear.
The depinning threshold remains zero as $\phi/\phi_J$ is reduced until
$N$ drops below $N_p$ and all the disks can be trapped, giving $F_c=F_p$. 
At high disk densities such as $\phi/\phi_{J} = 0.99$ in Fig.~1(b), 
$F_c$ is finite.
Figure 1(b) also shows an interesting crossing of the velocity-force curves, 
caused by the sudden jump in $\langle V_x\rangle/N$ at depinning for the
$\phi/\phi_J=0.99$ sample.
The depinning is elastic for $\phi/\phi_J=0.99$, so all the disks begin to move
simultaneously at the same velocity with only small localized
rearrangements of the
disk packing.
In contrast, for $\phi/\phi_J=0.947$ the depinning is plastic.
Some disks remain pinned while other disks move past, and the pinned disks do
not depin until a much higher $F_D$ is applied (not shown).
An example of the plastic flow occurring at depinning appears in
Fig.~1(d) for a sample with $N_p/N_J=0.415$ and $\phi/\phi_{J}=0.761$
at $F_{D} = 0.25$, above the depinning threshold.
Certain disks are always pinned while rivers of disks flow around them.   
In the jammed state, the disk packing acts like a solid and a small number of
pinning sites can trap all the disks, as in the sample with $N_p/N_J=0.09267$
and $\phi/\phi_J=0.99$ in Fig.~1(b).  The depinning from the jammed state is
elastic, as illustrated for $\phi/\phi_J=1.03878$ in Fig.~1(e,f).
Below jamming in samples containing a small number of pinning sites, 
such as the $N_p/N_J=0.09267$ and $\phi/\phi_J=0.947$ sample shown in 
Fig.~1(b), some
but not all of the disks can be immobilized, so the depinning is plastic. 

\begin{figure}
\includegraphics[width=3.5in]{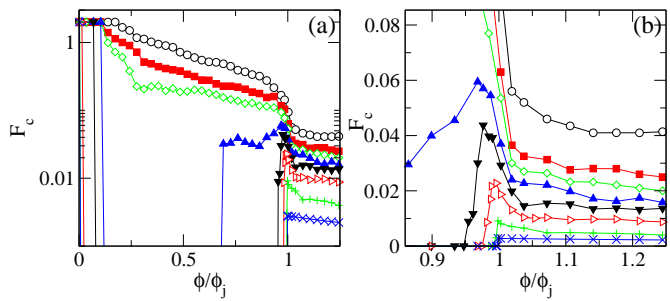}
\caption{
(Color online)
$F_{c} $ vs $\phi/\phi_{J}$ for 
$N_{p}/N_{J} = 0.828$ ($\bigcirc$), 
$0.415$ (red $\blacksquare$), 
$0.277$ (green $\lozenge$), 
$0.138$ (blue $\blacktriangle$), 
$0.09267$ ($\blacktriangledown$), 
$0.0346$ (red $\triangleright$), 
$0.00692$ (green $+$), 
and  $0.00138$ (blue $\times$).
(a) A log-linear plot over the full range of $\phi/\phi_J$
shows that the depinning threshold is always finite for 
$N_{p}/N_{J} > 0.138$, while for $N_p/N_J \leq 0.138$ there is a pinned
regime at low $\phi/\phi_J$ when all the disks are trapped by pinning
sites and another regime of finite $F_c$ at high $\phi/\phi_J$ where
jamming occurs.
(b) A blowup of the region near $\phi/\phi_{J} = 1.0$ 
shows that the onset of jamming is marked by either
a drop in $F_{c}$ for the higher pinning densities 
or a peak in $F_c$ for the lower pinning densities.  
}
\label{fig:2}
\end{figure}

Figure~2(a) shows $F_{c}$ versus $\phi/\phi_{J}$ 
for a series of different values of $N_{p}/N_{J}$, 
and Fig.~2(b) shows a blowup of the same data near $\phi/\phi_{J}  = 1.0$. 
For $N_{p}/N_{J} > 0.138$ the depinning threshold 
is finite for all $\phi/\phi_{J}$.
For $N_p/N_J \leq 0.138$
at low $\phi/\phi_{J}$, the depinning force 
reaches its maximum possible value of $F_{c} = F_p = 2.0$ since 
all the disks can be trapped by pinning sites without any overlap. 
There is a region of intermediate $\phi/\phi_J$
where there is no depinning threshold. 
At higher $\phi/\phi_J$, $F_c$ becomes finite again and the system
undergoes plastic depinning.  The value of $F_c$ increases with 
increasing $\phi/\phi_J$ in this regime until $F_c$ peaks
near $\phi/\phi_J=1.0$ when the system jams. 
At jamming, in principle a single pinning site could pin the entire
granular packing; however, we find that $F_c$ at jamming decreases as
the pinning density is reduced.
At and above jamming, the depinning becomes elastic 
and $F_c$ drops with increasing $\phi/\phi_J$ due to the increasing stiffness
of the jammed solid.

\begin{figure}
\includegraphics[width=3.5in]{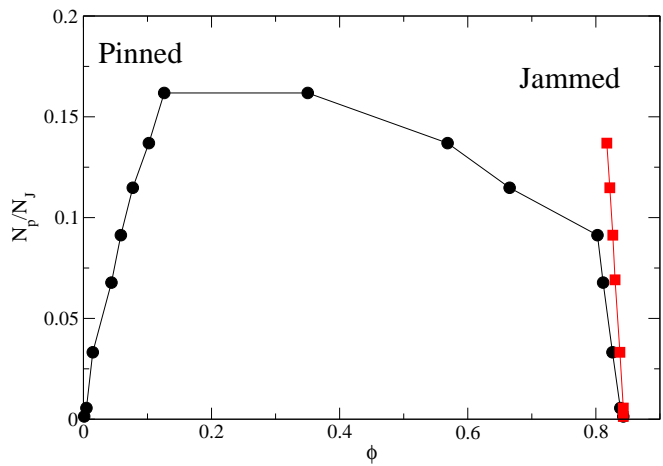}
\caption{ 
(Color online)
Black circles: The value of $N_p/N_J$ at which a finite depinning force
first appears vs $\phi$.
Red squares: The value of $N_p/N_J$ at which $F_{c}$ reaches its
peak value vs $\phi$.
The system depins elastically for $\phi$ just above the peak value of $F_{c}$. 
For small $\phi$, all of the particles are directly pinned at pinning
sites, while for $\phi$ close to $\phi_J=0.844$, the particles are pinned
due to jamming.
}
\label{fig:3}
\end{figure}
 
Using velocity force curves, we identify the onset of a finite depinning
threshold as a function of $\phi$ for different values of $N_p/N_J$, as well
as the value of $\phi$ at which $F_c$ reaches its peak value for $\phi$ 
near $\phi_J$.
The result is shown in Fig.~3.
There is a dome feature connecting the low $\phi$ limit, where the disks are
directly pinned at pinning sites, with the high $\phi$ limit, where the 
pinning occurs when the grains become jammed.
Near $\phi_J$ we find that the quenched disorder density $N_p/N_J$ can
be considered as a new axis of the jamming phase diagram, and that the
jamming density decreases with increasing pinning density.
Although the onset of jamming can be defined as coinciding with the appearance
of a finite depinning threshold, it could also
be defined as coinciding with the transition from plastic to elastic depinning,
where $F_c$ reaches its peak value.
Figure 3 shows that these two definitions are not identical but track each
other closely in the region near $\phi_J$.  
The onset of elastic depinning continues to produce a peak in $F_c$ 
up to the highest pinning densities we have considered.

We can make a simple argument for how quenched disorder reduces the
jamming density.
The average distance between pinning sites 
is $l_{p} \propto \rho_{p}^{-1/2}$, where $\rho_p=N_p/L^2$.
We can estimate a correlation
length $\xi$ by assuming that $\xi$ grows as jamming is approached according
to
$\xi \propto (\phi_{J} - \phi)^{-\nu}$. 
Jamming should occur when $l_p=\xi$, or when
$\rho_{p} \propto (\phi_{J} - \phi)^{2\nu}$.  A fit of
the onset of finite $F_c$ for $\phi > 0.8$ or a fit to the peak
value of $F_{c}$ both give 
a linear dependence on $\phi$, implying that $\nu = 0.5$.
This value is close to some threshold predictions 
\cite{Wyart}; however, caution must be taken
in comparing our exponent to systems without quenched disorder,
since the presence of quenched disorder or the fact that we are
driving our system could fundamentally change the
nature of the jamming compared to the disorder free case, 
and additional corrections
to scaling could be relevant \cite{Olsson}.  

\begin{figure}
\includegraphics[width=3.5in]{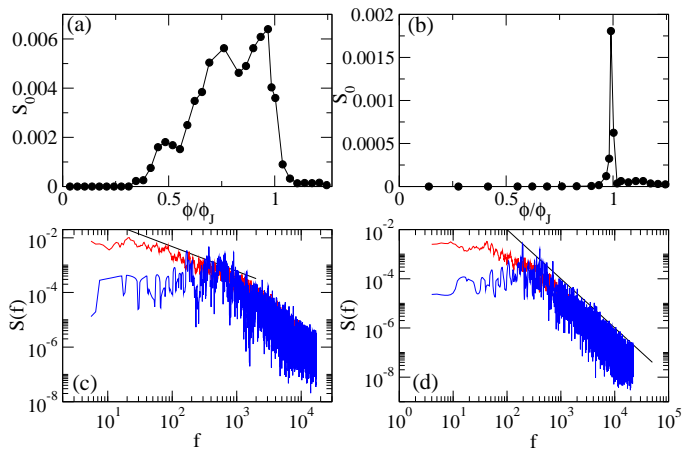}
\caption{
(Color online)
(a,b) The noise power $S_{0}$ obtained from the time series of the velocity
fluctuations vs $\phi/\phi_{J}$ 
for $F_D=1.1F_c$.
(a) $N_{p}/N_{J} = 0.277$. $S_0$ drops at the onset of 
jamming and at low $\phi/\phi_J$. 
(b) $N_{p}/N_{J} = 0.0346$. 
There is a pronounced peak in $S_0$ just below jamming.
(c) The power spectrum $S(f)$ vs $f$ from the system in 
(a) for (upper red curve) $\phi/\phi_{J} = 0.968$ where the 
depinning is plastic and (lower blue curve) $\phi/\phi_J=1.141$ where the
depinning is elastic and a narrow band noise signal appears.  The solid black
line is a fit to $1/f^{0.9}$. 
(d) $S(f)$ vs $f$ for the system in (b)
for (upper red curve) $\phi/\phi_{J} = 0.989$ and (lower blue curve) 
$\phi/\phi_{J} =1.141$ showing narrow band noise in the jammed phase.   
The solid black line is a fit to $1/f^2$.
}
\label{fig:4}
\end{figure}
 
The onset of jamming can also be detected by analyzing the
velocity noise fluctuations 
using the power spectrum $S(f)$ obtained from the time
series of the average disk velocities, 
$S(f)  = |\int \exp(-i2\pi ft)\langle V_x\rangle(t)dt|^2$. 
The noise power $S_{0}$ is defined to be the integrated noise power
in the lowest octave of the spectrum.
We take $F_D=1.1F_c$.
Figure~4(a) shows $S_{0}$ versus $\phi$ for $N_{p}/N_{J} = 0.277$ 
and Fig.~4(b) shows the same quantity for $N_{p}/N_{J} = 0.0346$.  
For $N_{p}/N_{J} = 0.277$, the noise power $S_0$ is low for 
$\phi/\phi_{J} < 0.3$ since there are very few
collective interactions at these low densities 
that could give rise to low frequency noise. 
$S_0$
decreases rapidly with increasing $\phi/\phi_J$ 
for $\phi/\phi_{J} > 0.95$ after the system jams and transitions to elastic
depinning.
At intermediate $\phi/\phi_J$, large velocity fluctuations occur and produce
a broad band noise signal,
as shown in Fig.~4(c) 
for $\phi/\phi_{J} =0.968$ where the depinning is plastic. 
The solid line in Fig.~4(c) is a fit to 
$1/f^{0.9}$. 
The appearance of $1/f$ type noise is known
to be associated with plastic depinning \cite{Marley,Olson}. 
For  $\phi/\phi_{J} = 1.141$ in the elastic depinning regime,
Fig.~4(c) shows that the noise power is considerably reduced 
and the spectrum has a peak at finite frequencies 
with several higher harmonics, indicative of a
narrow band noise signal.  
The appearance of narrow band noise in driven systems with quenched disorder is 
associated with the formation of a moving solid \cite{Gruner,Okuma} 
and is commonly referred to as a washboard frequency. 
This is consistent with the moving jammed packing acting like a 
rigid solid.
For the smaller pinning density of $N_{p}/N_{J} = 0.0346$, Fig.~4(b) 
shows a pronounced peak in $S_0$ at $\phi/\phi_{J} = 0.989$, 
which corresponds to a density just below the peak in $F_{c}$. 
$S_0$ then drops off rapidly as the system enters the jammed phase     
and depins elastically. 
In Fig.~4(d) we plot the power spectra for 
$\phi/\phi_J = 0.989$ and $\phi/\phi_J=1.141$ for $N_p/N_J=0.0346$.
There is broad band noise in the plastic flow regime at $\phi/\phi_J=0.989$ and 
narrow band noise in the elastic flow regime at $\phi/\phi_J=1.141$.  

The depinning-jamming transition we observe has many similarities to the 
peak effect phenomenon in type-II superconductors with vortices moving 
through random disorder. 
The peak effect can occur as a function of vortex density. 
At low densities,
the depinning threshold is high since vortices can be pinned individually. 
At intermediate densities, the depinning threshold remains at a low value until
some higher density is reached where there is 
a significant increase in the depinning threshold to a peak value.
This is associated with a large 
increase in the noise power 
\cite{Higgins,Marley,Banerjee}. 
The peak effect and the noise features become more prominent 
for cleaner samples with
less pinning \cite{Banerjee}. 
All these features are captured in our results.
The standard interpretation of the
peak effect is that it marks a transition 
from a weakly pinned solid to a more strongly pinned
disordered state. Our results suggest that the peak effect may 
be a general phenomenon in 
other systems
with quenched disorder close to some type of phase transition.  

In summary, we show how jamming behavior changes with the addition of 
quenched disorder 
using a simple model
of bidisperse disks that exhibit a well defined jamming density 
$\phi_{J}$ in the absence of quenched disorder. 
We propose that quenched disorder represents a new axis 
of the jamming phase diagram and that increasing quenched disorder density
decreases the disk density at which the system jams.
At low disorder densities,
the disk density at which a finite depinning threshold 
appears coincides with point J.  
There is also a reentrant finite depinning threshold at low disk densities
when all the disks are directly pinned.
We find a maximum in the depinning threshold 
just at the onset of jamming for low
disorder densities.   
When the disorder density is sufficiently large, the depinning threshold
is finite for all disk density values; 
however, proximity to $\phi_{J}$ produces clear effects in the form of
features in the velocity force curves as well as noise fluctuation signatures.
Below jamming, the depinning is characterized by plastic flow and $1/f$ noise 
characteristics 
with strong heterogeneities in the system,
while above jamming, the depinning is elastic with all the 
particles moving together and is characterized by a washboard noise. 
For high disorder density the onset of jamming  
is associated with a drop in the depinning threshold as 
opposed to the peak in depinning found at low disorder density. 
Our results show many similarities to the peak effect observed in 
high-temperature superconductors
where a peak in the depinning threshold occurs at both low and high 
vortex densities. 
Our results should be relevant for
systems exhibiting depinning transitions and jamming. 

This work was carried out under the auspices of the 
NNSA of the 
U.S. DoE
at 
LANL
under Contract No.
DE-AC52-06NA25396.
Z.N. was supported by NSF DMR-1106293 at WU.

\end{document}